\documentclass[english,aps,pra,showpacs,notitlepage,superscriptaddress,preprintnumbers,nofootinbib]{revtex4-1}
\global\long\def\order#1{\mathcal{O}\left(#1\right)}
\global\long\def\d{\mathrm{d}}

\def\za{Z\alpha}
\def\({\left(}  
\def\){\right)}

\usepackage{amsmath}
\usepackage{amssymb}
\usepackage{bm}
\usepackage{tikz}

\usepackage{slashed}
\usepackage{graphicx}
\usepackage{epsfig}

\usepackage{epstopdf}
\usepackage{esint}

\begin{document}

\title{Light-by-light scattering in the Lamb shift and the bound
  electron $g$ factor}

\author{Andrzej Czarnecki}
\author{Robert Szafron \footnote{Present address: Physik Department T31,
James-Franck-Str. 1, Technische Universit\"at M\"unchen,
D–85748 Garching, Germany}} 
\affiliation{Department of Physics, University of Alberta, Edmonton,
  Alberta, Canada T6G 2G7}
\preprint{Alberta Thy 27-16}
\preprint{TUM-HEP-1069/16}

\begin{abstract}
We compute an $\order{\alpha^2(Z\alpha)^6}$ contribution to the hydrogen-atom Lamb
shift arising from the light-by-light scattering. Analogous diagrams,
with one atomic electric field insertion replaced by an external magnetic
field, contribute to the gyromagnetic factor of the bound electron at
$\order{\alpha^2(Z\alpha)^4}$. We also calculate the contribution to the gyromagnetic factor from the muon magnetic loop. 
\end{abstract}
\maketitle

\section{Introduction}
Light-by-light scattering (LBL) arises when a virtual charged particle
induces an interaction among photons. Because of the charge conjugation symmetry of quantum
electrodynamics (QED), the number $n_\gamma$ of the coupled photons must be
even \cite{Furry:1937zz}.  When $n_\gamma=4$, a number of phenomena
result that manifest themselves, for example, by a slight change of
energy levels in hydrogen. Averaged over possible spin orientations of
the electron and the proton, this constitutes a part of the Lamb
shift. 

\begin{figure}[htb]
\noindent \begin{centering}
\includegraphics[scale=0.9]{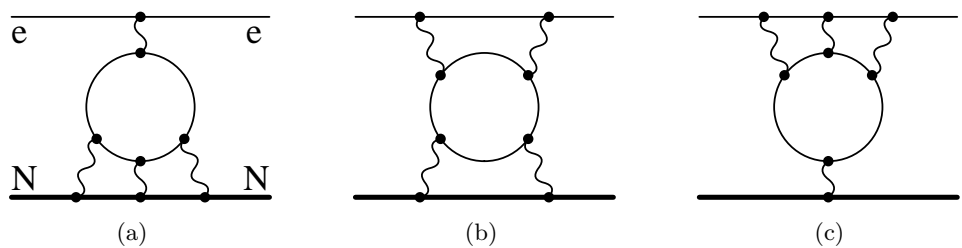}\\
\centering (a)  \hspace*{40mm} (b)  \hspace*{40mm} (c)
\par\end{centering}
\caption{\label{fig:LambLBL} Examples of light-by-light scattering contributions to
the Lamb shift. Thin lines denote electrons and the thick one is the nucleus.}
\end{figure}

Examples of possible processes are shown in Fig.~\ref{fig:LambLBL}. 
Fig.~\ref{fig:LambLBL}(a) shows the lowest order contribution of the
so-called Wichmann-Kroll potential, first considered in
\cite{Wichmann:1954zz,Wichmann:1956zz}. Its effect is
$\order{\alpha(\za)^6}$ and shifts the 1S level of hydrogen ($Z=1$)
by  2.5 kHz. For comparison, the total Lamb shift of the 1S level
starts at $\order{\alpha(\za)^4\ln\za}$ and is about 8 MHz (for a
review of the theory of the Lamb shift see \cite{eides2007theory,yerokhin2015lamb,Mohr:2015ccw}). 

Fig.~\ref{fig:LambLBL}(b) is an example of an
$\order{\alpha^2(\za)^5}$  effect. Such LBL effects were computed in
\cite{Eides:1994rc,Eides:1994dq,Pachucki:1993zz,Pachucki:1994ega} and confirmed in
\cite{Dowling:2009md}; they shift the 1S level by $-5.3$ kHz.  

Finally, Fig.~\ref{fig:LambLBL}(c) is an $\order{\alpha^3(\za)^4}$ LBL contribution to the Lamb shift
that enters through the slope of the Dirac form factor computed at
this order in \cite{Melnikov:1999xp}.

A peculiarity of atomic physics is that a given QED Feynman diagram gives
rise to contributions at various orders in $\za$. 
In this paper, we determine the effect of diagrams similar to
Fig.~\ref{fig:LambLBL}(b) in the next order in the $\za$ expansion,
$\order{\alpha^2(\za)^6}$. Effects related to the self-energy of the
electron at this order were studied in \cite{Jentschura:2005xu}, but
the LBL contribution was not included there. 

Of course, attaching an extra photon to the
electron loop would give zero because of the Furry theorem \cite{Furry:1937zz}. Instead,
we consider a different region of momenta $\bm{q}_{1,2}$ of the two photons exchanged
between the nucleus and the electron loop (``Coulomb photons'').

\begin{figure}[htb]
\newcommand{\spc}{52mm}
\includegraphics[scale=0.9]{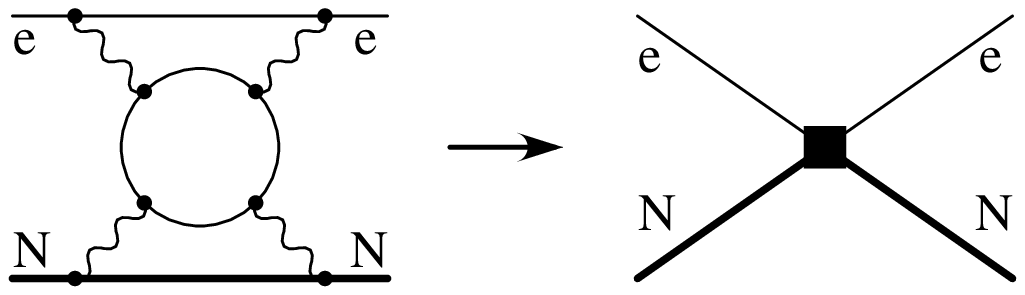}\\[1mm]
(a) \hspace*{\spc}(b) \\[3mm]
\includegraphics[scale=0.9]{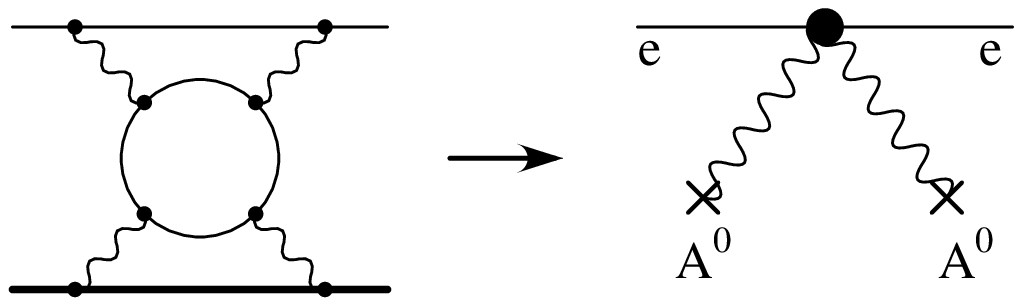}\\[1mm]
(c) \hspace*{\spc}(d) 
\caption{\label{fig:1}
Matching with all photons
hard (upper panel). Here, 
Coulomb photons are part of short-distance loops (a) that can
be matched onto a point interaction (b).  When the Coulomb field 
carries soft momentum (lower panel), the remaining two short-distance
loops in (c) are shrunk into an effective vertex connecting
two electron fields and two Coulomb photons (d). This results in
corrections $\order{(Z\alpha)^6}$.}  
\end{figure}

It is convenient to classify effects of photon exchanges according to
how the momentum they carry scales with the atomic number $Z$. 
Results $\order{\alpha^2(Z\alpha)^5}$ are obtained when the
Coulomb photons are hard, $\left|\bm{q}_i\right| \sim m_e \gg m_e Z\alpha$. 
They scale like the momentum inside the LBL loop that is of the
order of $m_e$ (there is no dependence on $Z$). In this case,
the momentum transferred between the electron and the nucleus (that
scales like $Z\alpha m_e$) can be neglected.
This gives rise to a contact interaction between the electron and the
nucleus, as shown in the upper panel of Fig.~\ref{fig:1}. This
interaction creates an  effective potential proportional to
$(\za)^2\delta^3(\bm{r})$  \cite{Pineda:1997ie, Hill:2012rh}, scaling with the nuclear charge as
$ \sim (\za)^5$ and contributing to the Lamb shift at $\order{\alpha^2
  (\za)^5}$. 

In this paper, we consider instead a situation where both Coulomb photons
carry a small momentum, $\vec{q}_i\ll m_e$. Then the LBL loop can be
expanded in $\bm{q}_i$. The leading term in this expansion is
proportional to
$(Z\alpha)^2\frac{\bm{q}_1 \cdot \bm{q}_2}{\bm{q}_1^2 \bm{q}_2^2}$
\cite{Jentschura:2005xu, Czarnecki:2005sz}. The resulting interaction is
illustrated in the lower panel of Fig.~\ref{fig:1}.  In position space
this term is proportional to the square of the electric field
$\bm{E}^2$. Since the electric field scales as
$\sim \za/r^2 \sim (\za)^3$, this contribution is
$\order{\alpha^2 (\za)^6}$.

The corresponding effect  on the Lamb shift is presented in
Section \ref{sec:Lamb}. An analogous effect on the bound-electron
gyromagnetic factor ($g$) is described in Section \ref{sec:g}.

\section{LBL contribution to the Lamb shift at $\order{\alpha^2(Z\alpha)^6}$}
\label{sec:Lamb}
\begin{figure}[htb]
\begin{tikzpicture}
    \begin{scope}
    \node { \includegraphics[width=0.5\textwidth]{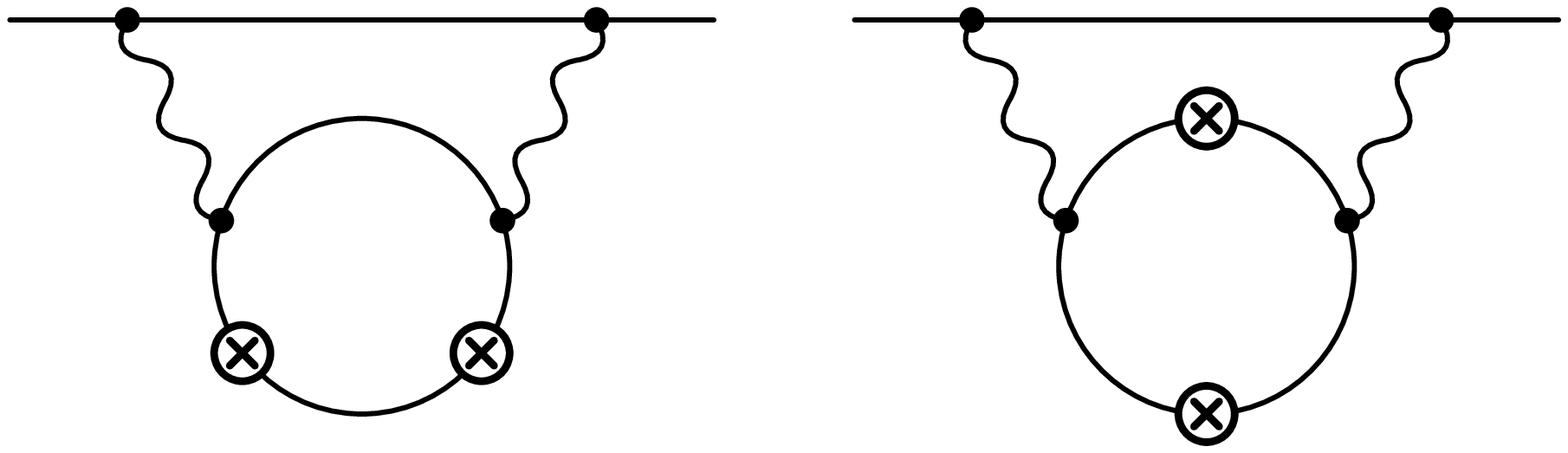} };
    \end{scope}
    \begin{scope}[xshift=74mm,yshift=-1.5mm]
    \node {\includegraphics[width=0.22\textwidth,]{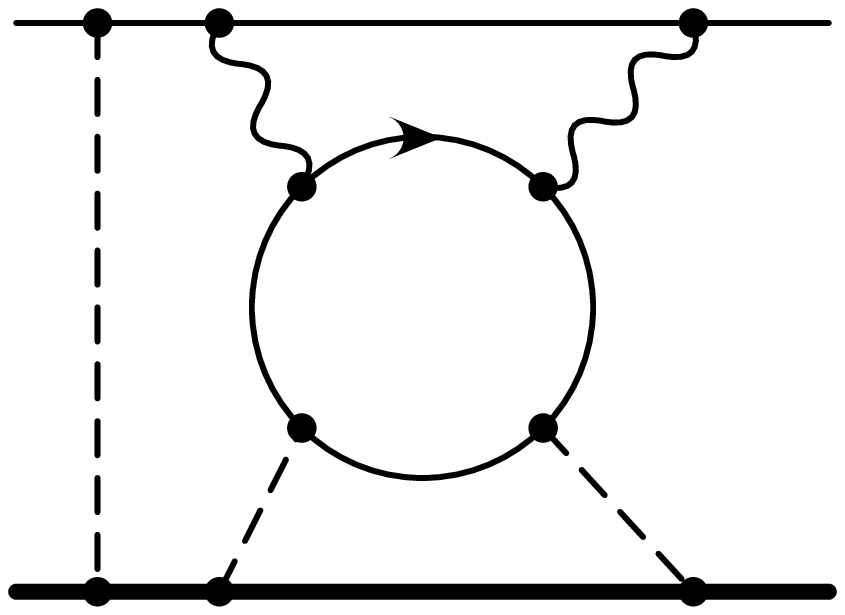}};
    \end{scope}
\end{tikzpicture}\\
 \hspace*{0mm}
(a)  \hspace*{41mm}
(b) \hspace*{44mm}
(c) 
\caption{\label{fig:LBL} Light-by-light scattering contributions to 
  the Lamb shift. Crosses in (a) and (b) denote couplings of the electric field photons in whose momenta we expand. (c): an example of a hard-momentum diagram which cancels the logarithmic divergence induced by (a) and (b). }
\end{figure}

The contribution of diagrams in Fig.~\ref{fig:LBL}(a,b) to the
scattering amplitude $T$ of an
electron on an electric field \cite{Jentschura:2005xu} is found by
computing two-loop integrals with the result
\begin{align}
  \label{eq:1}
  \Delta T_\text{LBL} &=  \chi_\text{LBL}\,\bm{q}_1\cdot \bm{q}_2,\\ \label{eq:chi}  
\chi_\text{LBL} &= \frac{43}{144} - \frac{133}{3456}\pi^2,
\end{align}
where $\bm{q}_{1,2}$ denote the momenta of the Coulomb photons in
Fig.~\ref{fig:1}(d). The effective operator induced by the diagrams in Fig.~\ref{fig:LBL} is proportional to the square of the electric field \cite{Jentschura:2005xu},
$\bm{E}^2\sim 1/r^4$. The expectation value of this operator 
in the hydrogen ground state has an ultraviolet divergence. In momentum space $r^{-4}\to k$, while
the Fourier transform of the charge density behaves at large $k$ as
$1/k^4$. Altogether, the expectation value behaves like $\int \frac{k
  \d^3 k}{k^4}$, and diverges logarithmically. 

This divergence is canceled by other diagrams, such as shown in
Fig.~\ref{fig:LBL}(c), where all photons are hard. In the sum of all
contributions only a logarithm of the ratio of scales survives. Its
contribution to the $n$S energy levels is
\begin{equation}
  \label{eq:2}
  \Delta E_n = \( \frac{\alpha}{\pi} \)^2 
 \frac{\(Z\alpha\)^6}{n^3}\ln \(Z\alpha\)^{2}\cdot 4 \chi_\text{LBL}. 
\end{equation}
For example, the 1S-2S energy splitting in hydrogen ($Z=1$) is decreased
by about 280 Hz.
For comparison, the experimental uncertainty is just 10 Hz
\cite{Parthey:2011lfa}. 

The 1S-2S splitting is also of experimental interest in the
hydrogen-like helium ion He$^+$  ($Z=2$) \cite{herrmann:2009aa,altmann:2016aa}. 
The correction we have found reduces that splitting by a much larger amount, 15.5 kHz.

The effect we have found modifies the so-called coefficient $B_{61}$  \cite{Pachucki:2001zz,Jentschura:2005xu} in front of the term $\order{\alpha^2(Z\alpha)^6\ln \(Z\alpha\)^{-2}}$.
The total linear logarithmic contribution to the  ground state  energy in this order becomes 
\begin{eqnarray}
  \Delta E_{1S} &=& \left(\frac{\alpha}{\pi}\right)^2 (Z\alpha)^6 \ln \(Z\alpha\)^{-2} \left[\frac{413\, 581}{64\, 800}+\frac{4}{3}N(1S)+\frac{2027}{864}\pi^2 -\frac{616}{135}\ln(2) \right.\\ \nonumber && \left. 
    -\frac{2}{3} \pi^2 \ln(2)+\frac{40}{9}\ln^2 (2)+\zeta (3) +\left(-\frac{43}{36}+\frac{133}{864}\pi^2 \right)_{\rm LBL} \right], 
\end{eqnarray}
where $N(1S)$ was calculated in \cite{Pachucki:2001zz}. 
The operator $\bm{E}^2$ contributes also to the normalized difference of expectation values for S states considered in  \cite{Jentschura:2005xu},
\begin{eqnarray}
  \left< \left< \frac{(\za)^2}{r^4}\right> \right>&=& n^3  \left<nS\left| \frac{(\za)^2}{r^4}\right| nS\right>- \left<1S\left| \frac{(\za)^2}{r^4}\right|1S\right>  \\
  &=&8(\za)^6 \left[ H_{n} -\ln n -\frac{2}{3} - \frac{1}{2n}+\frac{1}{6n^2} \right];
\end{eqnarray}
where $H_n=\sum_{k=1}^{n}\frac{1}{k}$ are  harmonic numbers.
Hence, this observable is also changed by our additional contribution
to the coeffiecient of $1/r^4$ in Eq.~(\ref{eq:chi}) (see Eq.~(2.10)
in  \cite{Jentschura:2005xu}). 
\section{The $g$ factor of a bound electron}
\label{sec:g}

\begin{figure}[htb]
\includegraphics[scale=0.9]{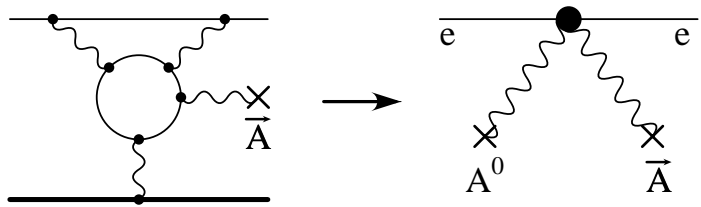}
\caption{\label{fig:2}
The LBL contribution to the bound electron $g$-factor is shrunk to a point interaction. The effective vertex contains two different 
operators, see text.}
\end{figure}

When one of the Coulomb photons in Fig.~\ref{fig:LBL}(a,b) is replaced
by an external magnetic field, we
obtain an effective interaction contributing to the bound electron 
$g$-factor, as shown in Fig.~\ref{fig:2}.
Momenta in both loops are on the order of the electron mass and the
loops can be treated as short-distance processes, compared with the
size of the atom. They induce effective low energy
operators, like in the case of the Lamb shift. 
We depict the matching procedure in Fig.~\ref{fig:2}. 
Two low-energy operators \cite{Pachucki:2005px}
contribute to the effective vertex,
\begin{equation}
\delta V =\frac{e^2}{2m}\left( 2\eta \sigma^{ij} B^{ik} \nabla^j E^k
  +\xi \sigma^{ij} B^{ij} \nabla^k E^k \right).
\label{deltaV}
\end{equation}
Here we use a $d$-dimensional notation ($d=3-2\epsilon$),
$\sigma^{ij}=\frac{1}{2i}\left[\sigma^i,\sigma^j\right]$ and $B^{ij}=\nabla^i
A^j - \nabla^j A^i $. The two terms in (\ref{deltaV}) differ only by the contraction of
vector indices. In an S-state only scalar averages of both operators
contribute. The $\order{\alpha^2}$ correction
to the $g$-factor due to the LBL contribution in an S-state is, in the limit $d\to 3$,
\begin{equation}
g^{(2)}_{\rm LBL} =  16(Z\alpha)^4 \left(\frac{2}{3}\eta +\xi  \right).
\end{equation}
We find that diagrams of the type shown in Fig.~\ref{fig:2} contribute 
$\Delta\eta =\left(\frac{\alpha}{4\pi}\right)^2 \left(\frac{31}{72} \pi ^2-\frac{22}{9}\right)$
and $\Delta\xi =\left(\frac{\alpha}{4\pi}\right)^2 \left(
\frac{16}{9}-\frac{25}{54} \pi ^2\right)$, giving
\begin{equation}
g^{(2)}_{\rm LBL} =(Z\alpha)^4 \left(\frac{\alpha}{\pi}\right)^2 \frac{16-19\pi^2}{108}.
\end{equation}
For completeness let us show the total contribution to parameters $\xi$ and $\eta$ at the order $\alpha^2$, including previously calculated vacuum polarization and self-energy diagrams \cite{Pachucki:2005px}
\begin{eqnarray}
 \eta &=&  \left(\frac{\alpha} {4\pi}\right)^2  \left[
\left( \frac{2528}{81} - \frac{169}{54}\pi^2\right)_{\rm VP}+ \left(\frac{31}{72} \pi ^2-\frac{22}{9}\right)_{\rm LBL} 
-\frac{283}{ 10}  + \frac{169}{120}\,\pi^2 - \frac{4 }{ 15}\,\pi^2 \ln 2 +
\frac{2}{5}\,\zeta(3) - \frac{16}{ 3 \varepsilon}\right], \\
\xi &=& 
 \left(\frac{\alpha}{ 4\pi}\right)^2 
\left[ \left(\frac{ 2674}{81}-\frac{91}{ 27}\pi^2\right)_{\rm VP} + \left(
\frac{16}{9}-\frac{25}{54} \pi ^2\right)_{\rm LBL}   -\frac{152}{15} + \frac{319}{ 45}\,\pi^2 -\frac{ 68}{5}\,\pi^2\ln 2
+\frac{102 }{ 5}\,\zeta(3)  + \frac{4 }{ 3  \varepsilon}
\right]\,.
 \end{eqnarray}
For the 1S state, the total correction to the
$g$-factor including the LBL contribution is 
\begin{equation}
g^{(2)} \approx (Z\alpha)^4 \left(\frac{\alpha}{\pi}\right)^2 \left[
  -18.03 - \frac{56}{9} \ln Z\alpha \right].
\end{equation}
The total correction of the order $\left(\frac{\alpha}{\pi}\right)^2 (Z\alpha)^4$, including the light-by-light contribution becomes
\begin{eqnarray}
g^{(2)} &=&
 \biggl(\frac{\alpha}{\pi}\biggr)^2\,\frac{(Z\,\alpha)^4}{n^3}\,\biggl\{
 \frac{28}{9}\,\ln[(Z\,\alpha)^{-2}]  +
   \frac{258917}{19440} - \frac{4}{9}\,\ln k_0 -
   \frac{8}{3}\,\ln k_3 + \frac{113}{810}\,{\pi }^2
  -  \frac{379}{90}\,{\pi }^2\,\ln 2
  + \frac{379}{60}\,\zeta(3)
 \nonumber\\  & +&\left(\frac{16-19\pi^2}{108}\right)_{\text{LBL}} + \frac{1}{n}\left[
   -\frac{985}{1728}   - \frac{5}{144}\,{\pi }^2 +
      \frac{5}{24}\,{\pi }^2\,\ln 2 - \frac{5}{16}\,\zeta(3)\right]
  \biggr\}\,,
\end{eqnarray}
where $k_0$ and $k_3$ are Bethe-logarithms defined and  calculated in \cite{Pachucki:2005px}. 

Measurements of the bound $g$-factor are the best current source of the electron atomic mass \cite{Sturm:2014bla}.
Since corrections $\order{\alpha^{2}(Z\alpha)^5}$ are not yet known, data with various values of $Z$ are used to fit them. In this approach, also the sensitivity to the presently found LBL effects is diminished. Once the  $\order{\alpha^{2}(Z\alpha)^5}$ corrections become available, an analogous fit will be used to constrain even higher order effects and further improve the knowledge of the electron mass. Then our $\order{\alpha^{2}(Z\alpha)^4}$ LBL result will allow for a reliable result.

A contribution to the bound-electron $g$-factor can also be obtained
from diagrams in  Fig.~\ref{fig:LBL}(a,b) by replacing one of the
photons connecting the electron line to the LBL loop by an external
magnetic field. This results in an effect $\order{\alpha
(Z\alpha)^5}$, already evaluated in \cite{Karshenboim:2002jc, Lee:2004vb}.
\section{Magnetic loop with virtual muons}
\label{sec:mu}

\begin{figure}[htb]
\includegraphics[scale=0.9]{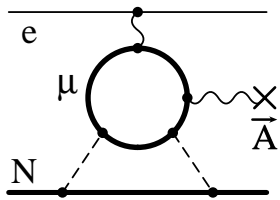}
\caption{\label{fig:mu} 
Magnetic loop with a virtual muon.}
\end{figure} 

Recently \cite{Belov:2016drs} the contribution of virtual muon loops
to the bound electron $g$-factor has been computed. Here we reevaluate
the magnetic loop, Fig.~\ref{fig:mu}, that on the basis of the results
in \cite{Belov:2016drs} seems easiest to confirm experimentally among
the LBL effects induced by virtual muons. We find its contribution
to be
\begin{eqnarray}
  \label{eq:3}
  g_\text{ML}(\text{muon}) = \frac{7}{216}\alpha\left(Z\alpha\right)^{5} \(\frac{m_e}{m_\mu}\)^3.
\end{eqnarray}
This turns out to be the same as the magnetic loop containing a
virtual electron  \cite{Karshenboim:2002jc,Lee:2004vb}, except for the additional factor
$\(m_e/m_\mu\)^3$. With hindsight this is easy to explain. 
The magnetic interaction in Fig.~\ref{fig:mu} involves a
scattering of the external magnetic field photon on the
virtual muon (that interacts with the nucleus via Coulomb photons
indicated by dashed lines) before coupling to the atomic
electron. This is an example of the Delbr\"uck scattering.
The amplitude of this scattering is inversely proportional to the cube
of the virtual fermion mass,  a consequence of  gauge invariance
\cite{Karshenboim:2002jc}.

Numerically, (\ref{eq:3}) gives $g_\text{ML}(\text{muon}) =5.5\cdot
10^{-22}$ for hydrogen, $Z=1$, and $4.3\cdot 10^{-18}$ for $Z=6$, the
much studied carbon ion. Both numbers are about two orders of
magnitude smaller than the values given in \cite{Belov:2016drs}. Thus,
contrary to the conclusion \cite{Belov:2016drs}, we believe that the
effect of the magnetic muon loop is too small to discern because of
nuclear uncertainties.
\section{Conclusions}
\label{sec:concl}

We have presented new contributions to the Lamb shift and the
bound-electron $g$ factor in hydrogen-like systems, arising from the
light-by-light scattering. 

For the $g$ factor, the new LBL contributions will influence the
determination of the electron mass when the
$\order{\alpha^{2}(Z\alpha)^5}$ corrections become available. We also
found that the effect of the muon magnetic loop is equal to the
analogous effect for the electron loop \cite{Karshenboim:2002jc}
multiplied by three powers of the electron to muon mass ratio. This
simple scaling is valid only for particles whose masses are larger
than the inverse atom radius and smaller than the inverse nucleus
radius. For light hydrogen-like ions, both the electron and the muon
satisfy these conditions but for example the tau lepton does not. For
the tau, we expect that the effect will be decreased by the nucleus
form-factor effects and further modified by the nuclear recoil.

For the Lamb shift, we have found a new logarithmic effect that decreases
the theoretical prediction for the 1S-2S splitting by an amount 28
times larger than the experimental error. This finding
strengthens the message of the recent review of the proton radius
puzzle \cite{Pohl:2013yb}: the theory of the hydrogen spectrum has to
be further scrutinized and its every aspect should be checked.

\begin{acknowledgments}
We thank Z.~Harman, R.~Pohl, and V.~A.~Yerokhin for useful comments. 
This research was supported by Natural Sciences and Engineering
Research Council (NSERC) of Canada.
\end{acknowledgments}



%

\end{document}